\documentclass{llncs}

\newif\ifdraft
\draftfalse

\usepackage[T1]{fontenc}
\usepackage[utf8]{inputenc}
\usepackage{lmodern}
\usepackage[scaled=.8]{beramono}
\usepackage{textcomp}

\ifdraft
\usepackage[show]{ed/ed}
\else
\usepackage[hide]{ed/ed}
\fi

\usepackage[hyperref=auto,firstinits=true]{biblatex}
\usepackage[babel]{csquotes}
\MakeAutoQuote{“}{”}
\MakeAutoQuote*{‘}{’}
\usepackage{paralist}
\usepackage{lstsemantic}
\usepackage{url}

\def\mysection#1{\section*{#1}}
\begin{document}

\title{Reimplementing the\\ Mathematical Subject Classification (MSC)\\ as a Linked Open Dataset\thanks{first author supported by DFG project I1-[OntoSpace] of SFB/TR 8 “Spatial Cognition” and EPSRC grant ``EP/J007498/1 – Formal representation and proof for cooperative games
    ''; second author by the University of Michigan.  
  The final publication is available at \texttt{http://www.springerlink.com}.}}
\author{%
Christoph Lange\inst{1,2,3}%
\and Patrick Ion\inst{4,5}%
\and Anastasia Dimou\inst{5}%
\and Charalampos Bratsas\inst{5}%
\and Joseph Corneli\inst{6}%
\and Wolfram Sperber\inst{7}%
\and Michael Kohlhase\inst{1}%
\and Ioannis Antoniou\inst{5}%
}
\institute{%
Computer Science, Jacobs University Bremen, Germany%
\and University of Bremen, Germany%
\and Computer Science, University of Birmingham%
\and American Mathematical Society%
\and Web Science, Aristotle University Thessaloniki, Greece%
\and Knowledge Media Institute, The Open University, UK%
\and FIZ Karlsruhe, Germany\\
Project page: \url{http://msc2010.org/mscwork/}}

\maketitle

\begin{abstract}
  The Mathematics Subject Classification (MSC) is a widely used scheme for classifying
  documents in mathematics by subject. Its traditional, idiosyncratic conceptualization
  and representation makes the scheme hard to maintain and requires custom implementations
  of search, query and annotation support. This limits uptake e.g. in semantic web
  technologies in general and the creation and exploration of connections between
  mathematics and related domains (e.g.\ science) in particular.

  This paper presents the new official implementation of the MSC2010 as a Linked Open
  Dataset, building on SKOS (Simple Knowledge Organization System).  We provide a brief
  overview of the dataset's structure, its available implementations, and first
  applications.
\end{abstract}

\mysection{Introduction}\label{sec:introduction}

The Mathematics Subject Classification (MSC~\cite{MSC2010}) maintained by Mathematical Reviews (MR) and Zentralblatt Math (Zbl) is a widely used system for classifying mathematical documents.  All major mathematical journals and digital libraries use the MSC\footnote{For details about this and about most other aspects of this project presentation, we refer to a full paper published recently~\cite{LIDBSKA:MSC-SKOS12}.}, mainly as a means of structuring literature in libraries and for the purposes of retrieving information by topic.  As the original format of the MSC2010 hindered its automated use and maintenance, we have reimplemented it as a machine-readable linked open dataset, which will soon be announced as the single official implementation.

\paragraph{MSC Usage and Maintenance So Far}

Previously, the right MSC class for a publication (e.g.\ 53A45 for “Vector and tensor analysis”) was typically chosen by consulting a human-readable document.  Web forms for uploading or searching publications typically required manual input of the desired MSC classes, rather than offering assistance.  The source of the MSC had been maintained in one plain \TeX{} file for almost 30 years
.  From this file, scripts produced several derived forms, including a KWIC index, a printable PDF, and HTML
.  Quality control of the source file with its custom macros could only be done by few experts, the derived forms did not particularly target machine processing, and the scripts that created them were once more custom, MSC-specific solutions.

\paragraph{Requirements for a Reimplementation}

After the \emph{content} of the MSC2010 had been settled, MR and Zbl decided to technically reimplement the source, according to the following requirements:
\begin{inparaenum}[(1)]
\item \textbf{facilitate reuse}, i.e.\ facilitate access, search, queries, and auto-classification; 
\item \textbf{facilitate maintenance}, i.e.\ preserve all of the original information, use a standard format supported by existing tools, enable integration of further maintenance information into the master source (such as the changes from MSC2000 to MSC2010);
\item \textbf{enable knowledge workers and service developers to adapt and extend} the MSC for their purposes, such as connecting mathematical subjects to related fields, adding descriptive labels in further languages, without affecting the core scheme; and
\item \textbf{allow end users to explore such connections} in order to \textbf{discover new knowledge}.
\end{inparaenum}
We chose to reimplement the MSC2010 as an RDF Linked Open Dataset, using the W3C-standardized SKOS vocabulary (Simple Knowledge Organization System~\cite{w3c:skos-reference}), which had been in use in digital libraries for several years.

\mysection{Structure of the MSC/SKOS Concept Scheme}\label{sec:design-mscsk-conc}

SKOS's built-in vocabulary covers a major part of the MSC concept scheme in a
straightforward way; however, for full MSC coverage, we had to extend SKOS in an
MSC-specific way.  We summarize the design below but refer to~\cite{LIDBSKA:MSC-SKOS12} for full technical detail.

\paragraph{Basics} The basic concept hierarchy was implemented as a SKOS concept scheme with 63 top-level concepts, which narrow down into 528 intermediate-level concepts, having 5606 final leaves.  SKOS also supports collections of concepts besides the main hierarchy, such as “all historical topics”.  The 5-character MSC class numbers (e.g.\ 53A45) are used as URIs of the concepts for identifying them and making their descriptions technically retrievable (see below).  SKOS allows for attaching multilingual labels to concepts; so far we have English, Chinese, Italian, and Russian, each from authoritative sources, and the RDF data model allows for maintaining further, non-authoritative translations separately.  Finally, SKOS supports links across concept schemes, which we have so far used for making explicit the changes from MSC1991 and MSC2000 to MSC2010.

\paragraph{Advanced Features} 

We go beyond the SKOS core, but follow established best practices, by linking MSC classes to concept schemes that have not yet fully been implemented in SKOS (the Dewey Decimal Classification).  We extended the SKOS vocabulary with partitive relations, i.e.\ when a link to a related concept is restricted to a certain scope such as “numeric approximations” or “applications in physics”.  0.4 percent of the concepts have labels containing mathematical markup, which cannot be expressed in Unicode but requires MathML.  RDF supports XML in labels, but it conflicts with multilinguality – a problem unsolved so far.  Finally, we attach some information, for which we have not yet designed a dedicated representation, such as co-classification policies, as generic notes to the respective concepts.

\mysection{Available Implementations}\label{sec:avail-impl}

\paragraph{The Dataset} We generated the new SKOS master source of the MSC2010 with a Perl script that translated the old \TeX{} source to RDF/XML.  We publish the data in four complementary ways, aiming to address a large audience of users and developers.  The project page at \url{http://msc2010.org/mscwork/} contains links to all alternatives.  For \textbf{linked data access} (i.e.\ directly retrieving an RDF description of each MSC class by dereferencing its URI; cf.~\cite{HB:LinkedData11}), we split the SKOS master file into one file per MSC class and serve the latter with the \texttt{application/rdf+xml} MIME type.  For \emph{querying} the dataset, we expose it through a SPARQL endpoint, i.e.\ a standardized interface for querying RDF.  For \emph{browsing}, we offer a wiki frontend.  Finally, for application developers, we offer the whole dataset for download as self-contained files in several formats.

Easy and reliable maintenance requires a master source without redundancy, e.g.\ the concept hierarchy should not be represented both top-down and bottom-up.  Linked data browsing, as well as other uses in large-scale applications that use plain RDF without inferencing support for performance, require a richer dataset with most practically relevant entailments expanded.  We automatically create this expanded version by applying a set of first-order rules implemented in N3, covering a subset of the SKOS semantics and our MSC-specific extensions.

\paragraph{In Use} The new implementation of the MSC2010 is currently in use on two websites linked from the project page.  Both have been developed independently using the RDF-aware Drupal 7 content management system.  On the homepage of the School of Mathematics at Aristotle University Thessaloniki (AUTH), we annotated the scientific fields covered by the courses and the faculty's research interests in terms of MSC/SKOS.  The new version of the PlanetMath encyclopedia (shortly entering beta testing) is powered by the Planetary system~\cite{KohDavGin:psewads11}.  In the old version, MSC-based navigation accounted for estimated 5–6 percent of all accesses.  We have now reimplemented this functionality in a more generic way: all article metadata (including MSC classifications) are maintained within \LaTeX\ preambles, the articles are transformed from \LaTeX\ to XHTML+RDFa, from which RDF is harvested into an RDF triple store, which also holds a copy of the MSC/SKOS dataset.  This approach is more flexible than the older static MSC access: our plan is to use SPARQL queries to expose specific slices of the encyclopedia content (e.g.\ “all articles by my co-authors in algebraic topology”).  The following listing, slightly adapted from the actual situation in PlanetMath, shows a SPARQL query that returns the subclasses of the given MSC class (which would actually be a parameterizable variable), and the number of articles classified with them\footnote{The number is determined using the \textit{COUNT()} aggregation function provided by the ARQ extension to SPARQL~\cite{ARQ}}:

\begin{lstlisting}[language=sparql,morekeywords={COUNT,GROUP,BY}]
PREFIX msc:  <http://msc2010.org/resources/MSC/2010/>
PREFIX skos: <http://www.w3.org/2004/02/skos/core#>
PREFIX dct:  <http://purl.org/dc/terms/>
SELECT DISTINCT ?subclass ?notation ?label COUNT(?article) WHERE {
  msc:53Axx skos:narrower ?subclass . # get subclasses; then, for each subclass:
  ?subclass skos:notation ?notation ;                 # get the MSC class number
            skos:prefLabel ?label .                    # get the preferred label
  OPTIONAL { ?article dct:subject ?subclass } # get classified articles (if any)
  FILTER langMatches(lang(?label), "en")                   # only English labels
}
GROUP BY ?subclass ?notation ?label   # grouping just needed for COUNT() to work
\end{lstlisting}

\mysection{Conclusion and Future Work}\label{sec:concl-future-work}

We deliver the first implementation of the MSC that is easily comprehensible to machines.  It does not only facilitate maintenance and development of novel services, but the rigorous conceptual modeling approach~\cite{LIDBSKA:MSC-SKOS12} also helped to uncover issues in the MSC conceptualization, which we have now fixed.  We are now planning to do three things in parallel: \begin{inparaenum}[(1)]\item refining the dataset implementation by making the internal structures of the MSC even more explicit\footnote{So far, our implementation makes two structural aspects more explicit than the old \TeX\ source: the concept hierarchy, which was previously given implicitly by the numbering scheme, and the cross-references, parts of which were previously given in natural language.  Our current implementation does not yet explicitly represent co-classification policies.} and by adopting further best practices for modeling classification schemes in SKOS~\cite{PanzerZeng:ClassificationSKOS}; \item supporting the MKM community in building applications that make use of the MSC, using our new implementation; and \item interlinking the MSC dataset with further mathematical and non-mathematical datasets, particularly including the OpenMath Content Dictionaries~\cite{Lange:KrextorSystem11} and DBpedia~\cite{dbpedia:on}.\end{inparaenum}  

\printbibliography
\end{document}